\documentclass[a4paper,11pt]{article}
\usepackage{pos}
\usepackage{booktabs}
\newcommand{\Mp}{M_{\rm P}}
\newcommand{\NGs}{\mathcal{N\!G}}

\title{Spiralling into Dark Energy: Axio-Dilaton Attractors in Light of Current Data}
\ShortTitle{Spiralling into dark energy}

\author*[a]{Mario Ramos-Hamud}

\affiliation[a]{DAMTP, University of Cambridge, Wilberforce Road, Cambridge, CB3 0WA, UK}

\emailAdd{mr895@cam.ac.uk}

\abstract{In string compactifications, dilaton-like moduli always come paired with axions, and the two fields interact through a curved field-space metric. We study the cosmology of such an axio-dilaton system, in which the dilaton rolls down an exponential potential corrected by a polynomial prefactor and couples to a massless axion through the kinetic function $W=e^{-\xi\phi_1/\Mp}$. A dynamical systems analysis shows that the late-time attractor follows a non-geodesic trajectory in field space with equation of state $\omega_\phi=(\lambda-2\xi)/(\lambda+2\xi)$, so accelerated expansion is possible for steep potentials provided $\xi\gg\lambda$. No physical fixed point is phantom. A Bayesian analysis with DESI DR2 BAO, binned Pantheon supernovae and Planck 2018 distance priors gives $\lambda\gtrsim7$--$9$ and $\log_{10}\xi\gtrsim2.5$ at 95\% confidence, together with a degeneracy between the two parameters. The model is slightly favoured over $\Lambda$CDM by current data, while the different polynomial corrections cannot yet be distinguished. This contribution is based on arXiv:2609.22539.}

\FullConference{
4th General Meeting of COST Action COSMIC WISPers (CA21106)\\
4th Training School of the COST Action COSMIC WSIPers (CA21106) (COSMIC-WISPers2026)\\
22-25 June and 16-20 June, 2026\\
Corfu, Greece\\}

\begin{document}
\maketitle

\section{Introduction}

Recent DESI BAO measurements \cite{DESI:2025zgx} have reopened the possibility that dark energy evolves in time. The standard way to model this is with a single scalar field and a potential chosen to fit the data. This approach has well-known limitations. The potential is not derived from an underlying theory. A single field has a flat target space, so it has no two-derivative interactions that could compete with gravity \cite{Brax:2023qyp,Smith:2025grk}. Exponential potentials, the natural outcome of string compactifications, do not easily give late-time acceleration in single-field models. Finally, string theory does not produce isolated light scalars: dilaton-like moduli, such as the overall volume or the inverse string coupling, always come with an axionic partner \cite{Cicoli:2023opf}, and neglecting the axion misses part of the low-energy dynamics.

These considerations single out the axio-dilaton system, which is also the basis of the Yoga dark energy proposal \cite{Burgess:2021obw}. Cicoli, Dibitetto and Pedro showed that the curvature of the field space allows accelerating solutions even for steep exponential potentials \cite{Cicoli:2020cfj}. In ref.~\cite{RamosHamud:2026}, written in collaboration with G.~Garc\'ia-Arroyo, F.~Quevedo and J.A.~V\'azquez, we extended this framework in two ways. We added the polynomial corrections to the potential that are expected from renormalisation-group effects, and we performed a Bayesian analysis of the model against current background data. This contribution summarises the main results. Derivations and the full stability analysis can be found in ref.~\cite{RamosHamud:2026}.

\section{The model}

We consider a dilaton $\phi_1$ and an axion $\phi_2$ with Lagrangian $\mathcal{L}/\sqrt{-g}=\tfrac12G_{ij}\partial_\mu\phi^i\partial^\mu\phi^j-V$ and field-space metric
\begin{equation}
ds^2=d\phi_1^2+W^2(\phi_1)\,d\phi_2^2\,,\qquad W(\phi_1)=e^{-\xi\phi_1/\Mp}\,,\quad\xi>0\,.
\end{equation}
As for axion--saxion pairs in supergravity, the axion direction is taken to be flat, so $V$ depends only on $\phi_1$. In the asymptotic region of moduli space the potential is exponential, $V\simeq V_0e^{-\lambda\phi_1/\Mp}$. We include the leading corrections through a polynomial prefactor,
\begin{equation}
V(\phi_1)=V_0\left[A+\left(\phi_1/\Mp-B\right)^{n}\right]e^{-\lambda\phi_1/\Mp}\,.
\label{eq:V}
\end{equation}
This form is not ad hoc. For a modulus $\tau$ the potential takes the form $f(\ln\tau)/\tau^m$, where $f$ comes from resumming the logarithms generated by renormalisation-group running: when particle masses scale with different powers of $\tau$, logarithms of mass ratios become logarithms of $\tau$ \cite{Burgess:2022nbx}. Since the canonical field satisfies $\phi_1\propto\ln\tau$, truncating $f$ gives eq.~\eqref{eq:V}. We consider $n=0$ (exponential, EXP), $n=1$ (Clemson--Liddle, CL \cite{Clemson:2008ua}) and $n=2$ (Albrecht--Skordis, AS \cite{Albrecht:1999rm}).

In a flat FLRW universe containing a barotropic fluid with $p_b=(\gamma-1)\rho_b$, the field equations read
\begin{equation}
\ddot\phi_1+3H\dot\phi_1-WW_{,\phi_1}\dot\phi_2^2+V_{,\phi_1}=0\,,\qquad
\ddot\phi_2+3H\dot\phi_2+2\frac{W_{,\phi_1}}{W}\dot\phi_1\dot\phi_2=0\,,
\end{equation}
with $3\Mp^2H^2=\tfrac12\dot\phi_1^2+\tfrac12W^2\dot\phi_2^2+V+\rho_b$. Since $W_{,\phi_1}<0$, the axion kinetic energy pushes the dilaton up its potential, against the force $-V_{,\phi_1}$. This competition is what allows acceleration on steep potentials.

\section{Dynamical systems analysis}

With $z_i=\phi_i/\Mp$, the variables $x_1=\dot z_1/(\sqrt6H)$, $x_2=W\dot z_2/(\sqrt6H)$, $y=\sqrt V/(\sqrt3H\Mp)$ and $\alpha=-V_{z_1}/V$ turn the equations of motion into
\begin{align}
x_1'&=-3x_1+\tfrac32\Pi x_1+\sqrt{\tfrac32}\left(\alpha y^2-2\xi x_2^2\right), &
x_2'&=-3x_2+\tfrac32\Pi x_2+\sqrt6\,\xi x_1x_2\,,\nonumber\\
y'&=\tfrac32\Pi y-\sqrt{\tfrac32}\,\alpha x_1y\,, &
\alpha'&=\sqrt6\,(1-\Gamma)\,\alpha^2x_1\,.
\label{eq:system}
\end{align}
Here a prime denotes $d/d\ln a$, $\Pi=2x_1^2+2x_2^2+\gamma\Omega_b$, $\Gamma=VV_{z_1z_1}/V_{z_1}^2$, and the Friedmann constraint reads $\Omega_\phi=x_1^2+x_2^2+y^2=1-\Omega_b$. The scalar equation of state is $\omega_\phi=(x_1^2+x_2^2-y^2)/\Omega_\phi$. The equation for $\alpha$ vanishes in three ways, each giving a different class of fixed points: (i) $x_1=0$, a frozen dilaton; (ii) $\alpha=0$, the dilaton sits at an extremum of $V$; (iii) $\Gamma=1$, where $V$ is locally a pure exponential.

\paragraph{Exponential potential ($n=0$).} Now $\alpha=\lambda$ is constant. Besides the single-field fixed points (kinetic domination $\mathcal{K}_\pm$, fluid domination $\mathcal{F}$, the scaling solution $\mathcal{S}$ and the geodesic point $\mathcal{G}$ \cite{Copeland:1997et}), the system has a pair of non-geodesic fixed points \cite{Cicoli:2020cfj},
\begin{equation}
\NGs_\pm:\quad x_1=\frac{\sqrt6}{\lambda+2\xi}\,,\quad x_2=\pm\frac{\sqrt{\lambda^2+2\lambda\xi-6}}{\lambda+2\xi}\,,\quad y=\sqrt{\frac{2\xi}{\lambda+2\xi}}\,,\qquad \omega_\phi=\frac{\lambda-2\xi}{\lambda+2\xi}\,,
\label{eq:NG}
\end{equation}
with $\Omega_\phi=1$. They exist and are stable for $\sqrt{6+\xi^2}-\xi\leq\lambda\leq2\xi$. Depending on the ratio $\lambda/\xi$, the equation of state interpolates between kination ($\lambda\gg\xi$), radiation ($\lambda=4\xi$), matter ($\lambda=2\xi$) and a cosmological constant ($\xi\gg\lambda$). The last limit holds for any $\lambda$: the curvature of the field space, not the flatness of the potential, is what drives acceleration. Because $\lambda>0$, the equation of state never crosses $-1$. A realistic history requires $\Omega_\phi\simeq0.7$ today, so the universe must be on its way from matter domination towards $\NGs_\pm$ rather than at the fixed point itself.

\paragraph{Clemson--Liddle potential ($n=1$).} After redefining $V_0$, $V=V_0(1+Cz_1)e^{-\lambda z_1}$ with $C=A-B$, and $\alpha'=\sqrt6(\lambda-\alpha)^2x_1$. The maximum of the potential gives a new de Sitter point $\hat{\mathcal{P}}$, which is always a saddle. All other fixed points coincide with those of the exponential case, and $\alpha$ approaches $\lambda$ as a power law, $\lambda-\alpha\sim N^{-1}$. The linear correction therefore leaves the late-time dynamics unchanged.

\begin{figure}[t]
\centering
\includegraphics[width=0.49\textwidth,trim=55 70 8 115,clip]{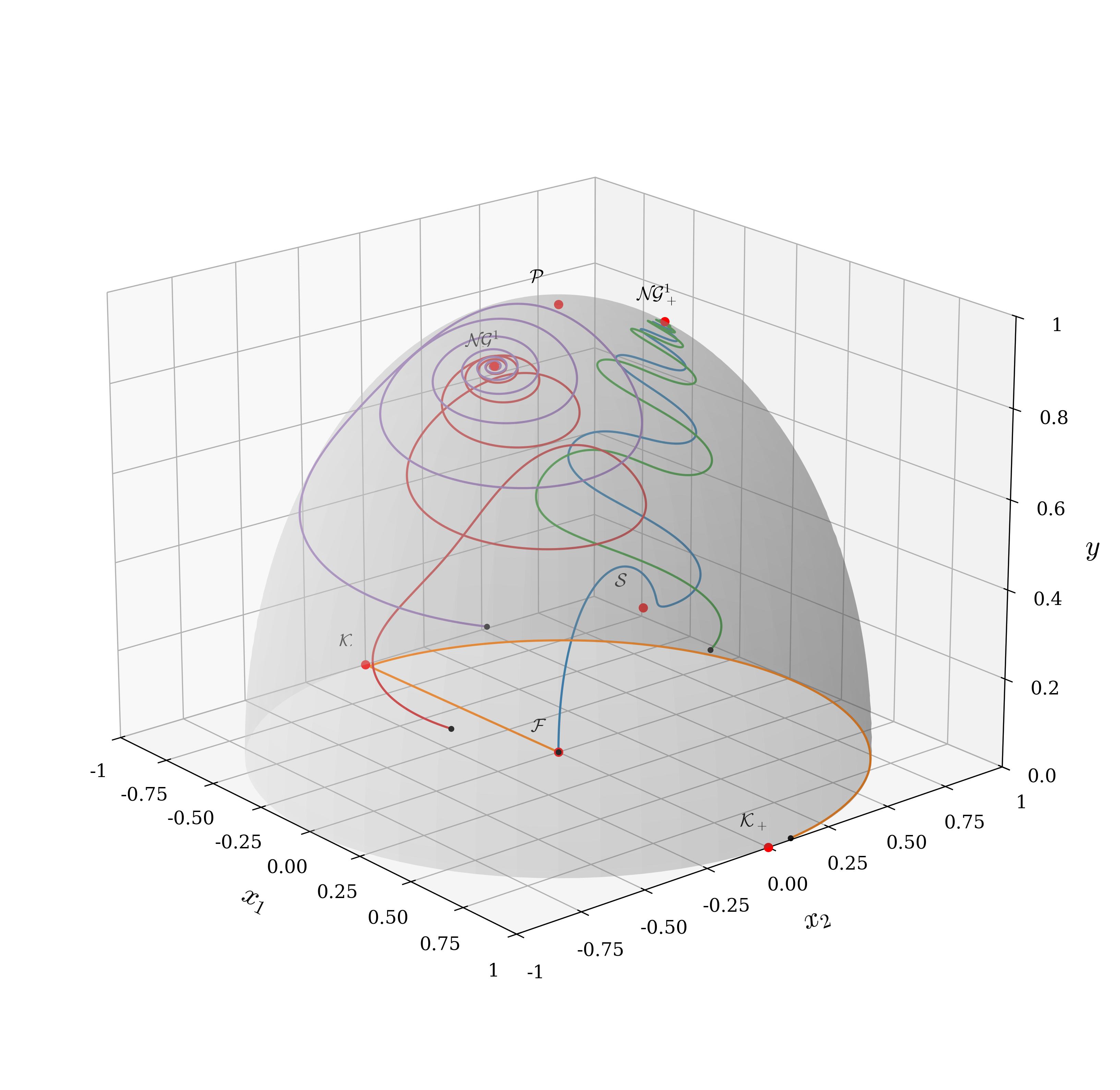}\hfill
\includegraphics[width=0.49\textwidth,trim=55 70 8 115,clip]{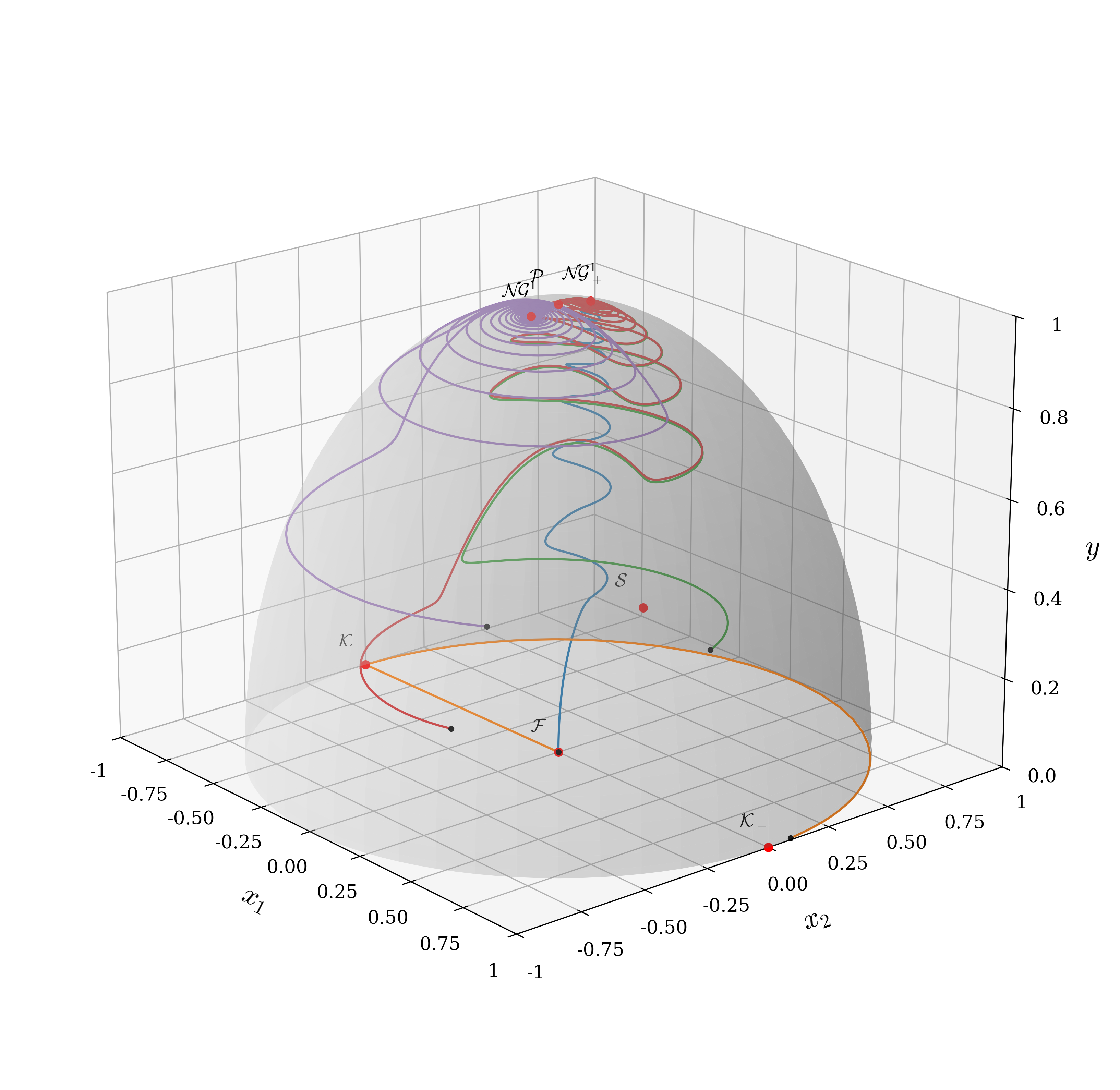}
\caption{Phase-space trajectories for the AS potential with $\gamma=1$, $\lambda=2$, $A=1$, and $\xi=10$ (left) or $\xi=100$ (right). Coloured lines are trajectories with different initial conditions (black dots); red dots are fixed points. The non-geodesic points act as late-time attractors. From ref.~\cite{RamosHamud:2026}.}
\label{fig:phase}
\end{figure}

\paragraph{Albrecht--Skordis potential ($n=2$).} The potential has a minimum and a maximum only when $A\lambda^2\leq1$. The minimum gives a de Sitter point $\mathcal{P}=(x_1,x_2,y,\alpha)=(0,0,1,0)$, which is a stable attractor for $A\lambda^2<1$ independently of $\lambda$ and $\xi$. The condition $\Gamma=1$ holds at $z_1-B=\pm\sqrt A$, where the potential behaves locally as an exponential with slopes
\begin{equation}
\alpha_1=\lambda-\frac{1}{\sqrt A}\,,\qquad\alpha_2=\lambda+\frac{1}{\sqrt A}\,.
\end{equation}
The fixed points of the exponential case reappear with $\lambda\to\alpha_{1,2}$, including two non-geodesic pairs $\NGs^{1,2}_\pm$ with $\omega_\phi=(\alpha_{1,2}-2\xi)/(\alpha_{1,2}+2\xi)$. The point $\NGs^1$ is physical only for $A\lambda^2>1$, since for $A\lambda^2\leq1$ one has $\alpha_1\leq0$ and the fixed point leaves the physical phase space. The line $A\lambda^2=1$ therefore separates two regimes. For $A\lambda^2<1$ the universe can end at the de Sitter minimum $\mathcal{P}$, which may coexist with $\NGs^2$. For $A\lambda^2>1$ the minimum disappears and the late-time attractor is $\NGs^1$. Again, no physical fixed point is phantom: $\omega_\phi<-1$ would require $-2\xi<\alpha<0$, where $x_2$ is imaginary. Figure~\ref{fig:phase} shows trajectories in the AS phase space.

\section{Bayesian analysis}

We performed a Bayesian analysis with \texttt{SimpleMC} \cite{simplemc,Aubourg:2014yra,Padilla:2019mgi}, combining DESI DR2 BAO \cite{DESI:2025zgx}, the binned Pantheon supernova sample (BP) \cite{Pan-STARRS1:2017jku} and the Planck 2018 distance priors (PLK) \cite{Chen:2018dbv}. We varied $\Omega_m$, $\Omega_bh^2$ and $h$ together with $\lambda\in[1,20]$, $\log_{10}\xi\in[1,3]$ and $\log_{10}A\in[-3,0]$, all with flat priors. We fixed $n=0,1,2$ in separate runs and, in a fourth run, varied $n\in[0,2]$. Table~\ref{tab:bayes} summarises the results.

\begin{table}[t]
\centering
\begin{tabular}{lccccc}
\toprule
 & $n=0$ & $n=1$ & $n=2$ & $n$ free & $\Lambda$CDM\\
\midrule
$\lambda$ & $>6.90$ & $>8.57$ & $>9.05$ & $>8.50$ & --\\
$\log_{10}\xi$ & $>2.57$ & $>2.51$ & $>2.55$ & $>2.49$ & --\\
$\chi^2_{\rm min}$ (DESI BAO) & 5.8 & 6.2 & 4.2 & 5.6 & 10.28\\
$\chi^2_{\rm min}$ (DESI BAO + BP + PLK) & 50.4 & 50.4 & 50.4 & 50.6 & 51.6\\
\bottomrule
\end{tabular}
\caption{95\% lower bounds on the scalar-sector parameters from DESI BAO + BP + PLK, and minimum $\chi^2$ for both data combinations \cite{RamosHamud:2026}. The parameters $A$ and $n$ are not constrained.}
\label{tab:bayes}
\end{table}

DESI BAO alone has little constraining power on the scalar sector. Adding BP and PLK, a clear correlation between $\lambda$ and $\xi$ appears for every $n$: a steeper potential needs a stronger kinetic coupling to keep $\xi\gg\lambda$ and hence late-time acceleration, as expected from the fixed-point analysis. The lower bound on $\lambda$ comes from early times, since a smaller $\lambda$ gives a larger early dark energy fraction. For $n=1$ the constant $A$ can be absorbed into a shift of the field, and the constraints are close to the exponential case. For $n=2$ the posterior of $A$ remains broad, while the effective slopes $\alpha_{1,2}$ are better constrained, so the data are sensitive to the combinations that enter the dynamics rather than to the shape of the correction. When $n$ is free, its posterior covers the whole prior range.

The background expansion reconstructed from the posteriors is very similar for all $n$. The Hubble rate stays close to the $\Lambda$CDM prediction, while the dark energy equation of state is close to $-1$ at early and late times, with a transient departure at $2\lesssim1+z\lesssim50$. The axio-dilaton models reach a lower $\chi^2_{\rm min}$ than $\Lambda$CDM for both data combinations, from 10.28 to 4.2--6.2 for DESI BAO alone and from 51.6 to about 50.4 when BP and PLK are included. The data therefore show a slight preference for the axio-dilaton models. This preference is modest: the models have more free parameters than $\Lambda$CDM, and the improvement shrinks considerably once all data sets are combined.

\section{Conclusions}

A curved field space is enough to obtain dark energy from the steep exponential potentials expected in string compactifications. The late-time attractor follows a non-geodesic trajectory with $\omega_\phi\to-1$ for $\xi\gg\lambda$, and no physical fixed point is phantom. Polynomial corrections from renormalisation-group running shift the effective slope and, for the AS potential with $A\lambda^2<1$, add a stable de Sitter minimum. Current background data favour steep potentials with large couplings, $\lambda\gtrsim7$--$9$ and $\log_{10}\xi\gtrsim2.5$, and show a slight preference for the model over $\Lambda$CDM, although they do not yet select a value of $n$. Perturbations and growth data, an axion potential, and interactions with dark matter, which could produce an effective phantom regime \cite{Smith:2025axd}, are natural next steps.

\acknowledgments
We would like to thank Gabriela Garc\'ia-Arroyo, Fernando Quevedo and J. Alberto V\'azquez for the collaboration on this project. This article is based upon work from COST Action COSMIC WISPers CA21106, supported by COST (European Cooperation in Science and Technology).

\end{document}